\documentclass[letterpaper, 10 pt, conference]{ieeeconf}      
\IEEEoverridecommandlockouts                              
\usepackage[pdftex]{graphics} 
\usepackage{epsfig} 
\usepackage{mathptmx} 
\usepackage{amsmath} 
\usepackage{amssymb}  

\title{
Fuzzy Logic based Autonomous Parking Systems - Part II:\\ A Hybrid Dual Controller System}

\author{Yu Wang and Xiaoxi Zhu
\thanks{The corresponding author Yu Wang is with the Department of Electrical Engineering, Yale University, New Haven, CT, 06511
        {\tt\small yu.wang@yale.edu}, Xiaoxi Zhu is with Singapore Telecommunications Limited, Singapore {\tt\small zhu.xiaoxi@hotmail.com}}}

\begin{document}

\maketitle
\thispagestyle{empty}
\pagestyle{empty}

\begin{abstract}

This paper presents an intelligent autonomous parking system with Hybrid Fuzzy Controllers (HFCs). The system enables intelligent vehicles to perform slot detection, parallel and vertical parking in a completely unmanned environment. The HFC, constituting of a Base Fuzzy Controller (BFC) and a Supervisory Fuzzy Controller (SFC), optimizes the control logic to counteract external disturbances in parking process by implementing additional fuzzy rule base. Customized HFCs are designed for critical steps in parking, namely turning control and posture stabilization. As a result, more accurate and efficient parking is achieved even when there are uncertainties in vehicle length and friction. Simulated experiments are carried out in MATLAB to verify the robustness of new HFCs and to demonstrate the performance improvement compared with the previous Fuzzy-Based Onboard System (FBOS). 

$\it Key words$--Autonomous Parking, Intelligent System, Adaptive Fuzzy Control, Parallel Parking, Vertical Parking, Driver's Assistance System, Intelligent Transportation System (ITS)

\end{abstract}

\section{INTRODUCTION}

The general topic of Intelligent Transportation System (ITS) has been widely explored in search of solutions to transportation problems. Some examples of the the popular research directions include driver's assistance system for safe driving and unmanned autonomous parking. Extensive research has been done in these areas, however, with limited progress. One major challenge comes from the fact that there are too many uncertainties in driving such that an accurate analytical model of the system is difficult to obtain. Even with an advanced controller which is able to capture the complexity of driving process, the  high computational cost of real-time processing hinders the wide application of these research findings. A big breakthrough in this area is Artificial Intelligence (AI) techniques which aim at a human-like driving control. One of the most widely applied techniques is fuzzy logic control.

First proposed by L.A. Zadeh, fuzzy logic control makes control decision in a way that resembles human reasoning ([1],[2]). A Fuzzy Logic Controller (FLC) does not require extensive knowledge of the process, but yet achieves satisfactory results. The superior performance comes from the robustness of underlying fuzzy logic. Fuzzy logic control is commonly applied in speed control and autonomous parking. Early development in speed control is focused on conventional Cruise Control (CC), where FLC controls the accelerator to maintain a constant speed ([3],[4]). Adaptive Cruise Control (ACC) takes one step further to keep a safety gap between two vehicles in the same lane ([4],[5]). Researches in FLC application to autonomous parking can be grouped into two general categories, tracking method ([6],[7]) and posture stabilization method ([8],[9]). Tracking method focuses on control algorithm to define optimal trajectory, while posture stabilization aims at achieving optimal final posture regardless of its initial status. One major drawback of both approaches is the high computational cost which prohibits the wide application in industry. 

The work presented in this paper is an intelligent auto-parking system with hybrid fuzzy logic controller as an effective solution to parking problems. Unmanned vehicles equipped with the system are able to achieve slot detection and auto-parking in either parallel or vertical parking mode. Fundamental control logics, including speed control, turning angle control and posture stabilization, are designed and implemented to ensure accuracy and efficiency in the parking process. The system is divided into different functional blocks (auto-driving, slot detection, parallel parking and vertical parking), each supported by a combination of different control logics. Section II gives an overview of the Hybrid Fuzzy Controller (HFC) by comparing it with previous Fuzzy-Based Onboard System (FBOS). The detailed HFC design is discussed in Section III and IV. Simulation and real-model testing are carried out and the results are summarized in Section V. Section VI briefly discusses the directions of future improvement.

\section{Hybrid Fuzzy Controller}

The Hybrid Fuzzy Controller (HFC) presented in this paper is an improvement of the Fuzzy-Based Onboard System (FBOS)  designed by the same research group [10]. The basic functionalities remain the same. Vehicles equipped with HFC or FBOS are able to achieve autonomous parking without human intervention. Currently, most of the driving assistance systems in market require drivers to find a suitable parking slot. As its major advantage, FBOS integrates two functions, parking slot detection and autonomous parking, into one single system. Moreover, different parking modes, parallel parking or vertical parking, can be automatically selected. 

The flow of autonomous parking under the control of FBOS is re-captured here. Driver leaves the car at car-park entrance and switches it to auto-parking mode. The car proceeds into the car park and starts searching for available slots under searching mode. Once a suitable slot is detected, either parallel-parking or vertical-parking mode is activated. An exit mode is also implemented which enables the car to move out of the parking slot and to drive to car-park exit. 

In the entire process, FBOS makes control decisions by analysing measurements from different sensors, which include angular/linear velocities and distance. In the 1:14 mini-scale vehicle prototype, measurements are gathered from infrared sensors and IMU. Based on the data, three important control tasks are performed, thus posture stabilization, turning control and parking slot detection. The original FBOS delivered accurate slot detection and smooth parking process in testing. However, the testing was carried out in a less dynamic environment, where same slot dimension, fixed vehicle size and uniform ground condition were maintained. In real-life situations, there are more complications which tend to degrade controller performance. Among all disturbances, friction (between tires and ground) and vehicle length are the two most important ones. If there are large deviations in these parameters, the outcome of autonomous parking might not be as perfect. 

The fuzzy logic controller is therefore re-designed to overcome the limitations discussed above. The new Hybrid Fuzzy Controller (HFC) significantly increases the system robustness. It allows vehicles with different lengths to park properly and smoothly regardless of ground conditions. Major improvements are made in turning control and posture stabilization. Section III and IV discuss the design of HFCs in detail.


\section{Turning Control}

The most crucial step in autonomous parking is to turn the vehicle around by a fixed angle. Take vertical parking as an example, the vehicle needs to turn around 90 degrees to fit into the parking slot (assuming that the vehicle has already adjusted its position properly before parking). The accuracy of this step is critical for a successful parking, especially when the size of parking slot is limited. 

Previous steering angle controller based on fuzzy logic demonstrated good results. However, the controller performance is degraded if there is large deviation in system parameters. An intuitive example is the vehicle length. A longer vehicle normally undergoes a larger turning radius than shorter ones. Therefore, if the steering-angle configuration is optimized for a shorter vehicle, it will not work on a longer one. The other major concern is the road friction. With coarse ground and new tires, the friction is noticeably larger and the vehicle tends to move much slower. The parking process is not smooth and takes longer time. On the contrary, with wet floor and worn tires, the result will be fast movement or even slipping. The turning trajectory will deviate from desired path in the above cases. 

The improved HFC for turning control consists of two separate control paths. One is to control the steering angle of the front wheels. This is to ensure that the vehicle follows a fixed trajectory during turning regardless of its own length. The other control path manages the speed during turning so that there is minimal jittering or slipping when the friction varies. Each of the paths consists of two fuzzy logic controllers. One is the Base Fuzzy Controller (BFC) that controls the steering angle and speed. The other is the Supervisory Fuzzy Controller (SFC) that fine-tunes the control signals. 

The design of BFC remains the same as in FBOS. The vehicle is equipped with IMU for angle and speed measurement. Before turning starts, angle measurement is cleared to zero. The set point is the desired angle to be turned (e.g. $90^\circ$ in vertical parking). In the middle of turning, velocity (linear and/or angular) and current angle is continuously monitored. Based on the control law, both front-wheel steering angle and speed should be large at the start and gradually reduces as the vehicle approaches the pre-set target.

\subsection{Steering Angle Control}

   \begin{figure}[thpb]
     \centering
     \includegraphics[width=7cm,height=4cm]{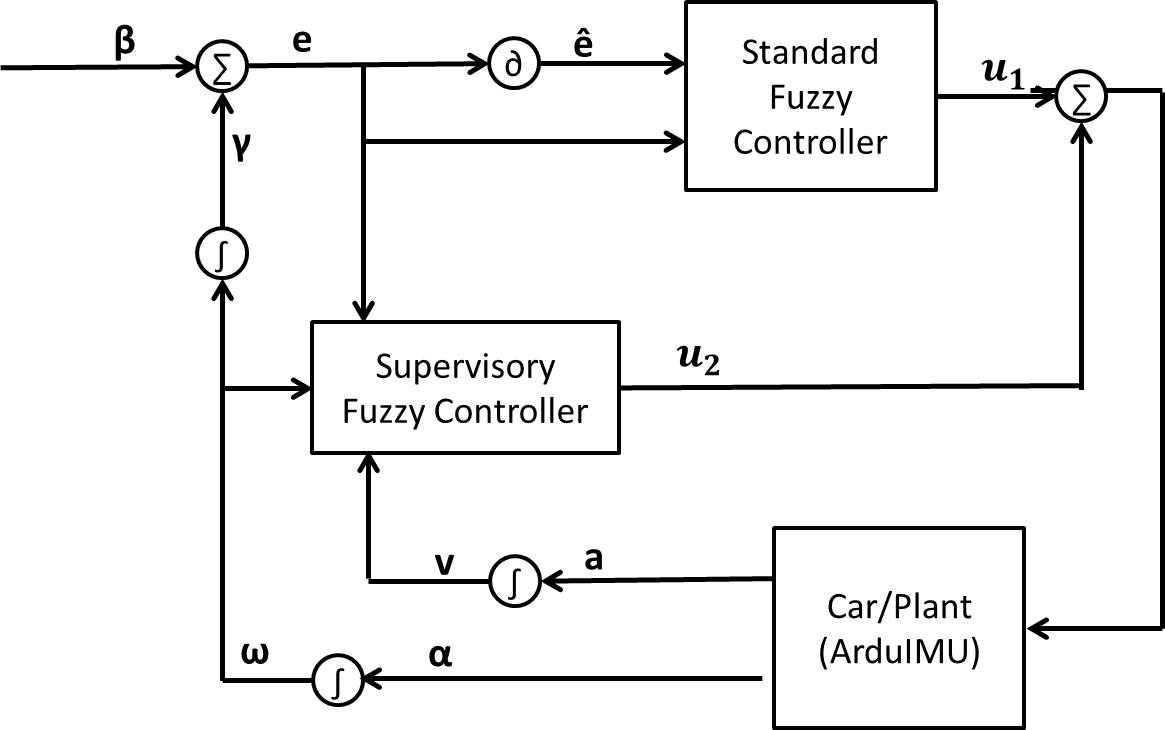}
     \caption{Steering Angle Control}
     \label{figurelabel}
   \end{figure}

Fig.1 illustrates the block diagram for the control path of front-wheel steering angle.\footnote{Symbols in the control diagram: $\beta$ is the set angle, a is the the linear acceleration and v is the linear velocity; $\alpha$ is the angular acceleration, $\omega$ is the angular velocity and $\gamma$ is the accumulated angle.} Inputs to the BFC are the difference between set angle and current angle (e), and change rate in that difference ($\hat{e}$). The two inputs are fuzzified based on the membership functions in Fig.2. The output is a voltage signal ($u_1$) sent to the servo, controlling the steering angle. The larger the voltage signal is, the larger the steering angle. Positive voltage (i.e. positive direction) indicates turning in the clockwise direction and vice versa. The IF-THEN rules for the BFC are summarized in Table I. 


   \begin{figure}[thpb]
     \centering
     \includegraphics[width=5cm,height=2.5cm]{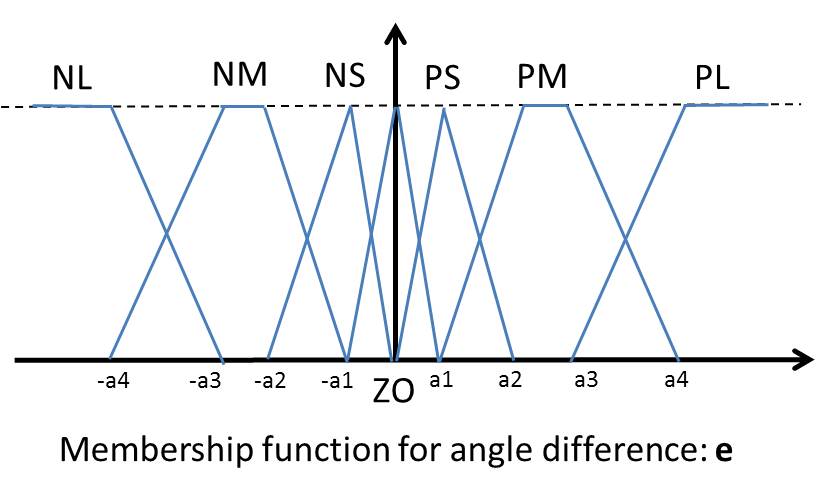}
	\end{figure}

	\begin{figure}[thpb]     
     \centering$
\begin{array}{cc}
    \includegraphics[width=2.7cm,height=1.8cm]{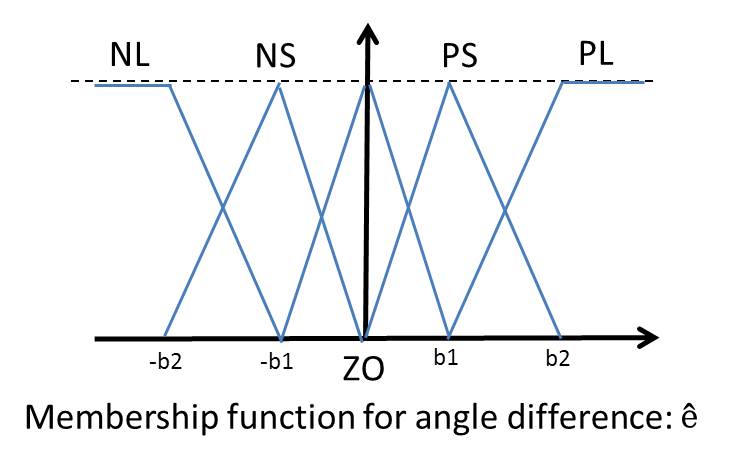}&
    \includegraphics[width=4.5cm,height=2cm]{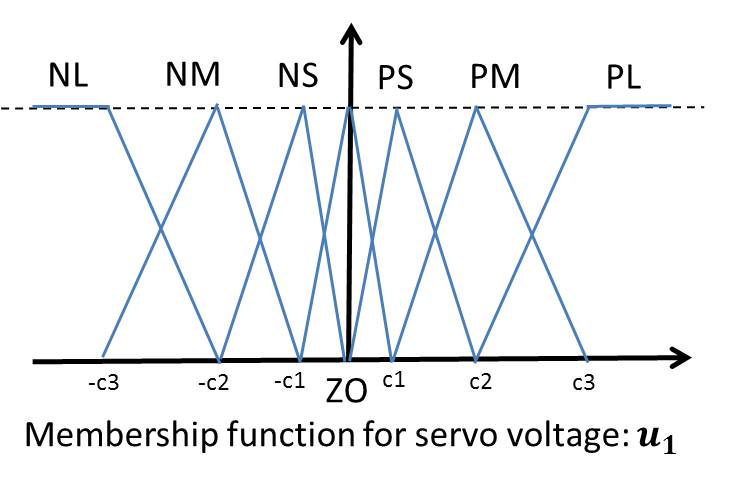}
\end{array}$
     \caption{BFC Inputs/Output in Steering Angle Control}    
   \end{figure}

The BFC gives satisfactory performance for normal-sized vehicles. Path {\bf a} in Fig.3 illustrates the trajectory of such a vehicle turning $90^\circ$. However, deviation occurs if there is a significant variation in vehicle length. Consider the extreme case when a mini cooper and a limo are trying to fit into the same vertical parking slot. The limo will follow path {\bf c} while mini cooper follows path {\bf a} instead. In spite of the different trajectories, it is also noticed that for each path the turning radius is smaller at the beginning and larger in the end. Therefore, a Supervisory Fuzzy Controller (SFC) is designed to achieve a uniform turning trajectory (path {\bf b} as in Fig.3).


	\begin{table}
	  \caption{IF-THEN Rules for BFC in Steering Angle Control}
  	  \begin{center}
   	  \includegraphics[width=6.5cm,height=3cm]{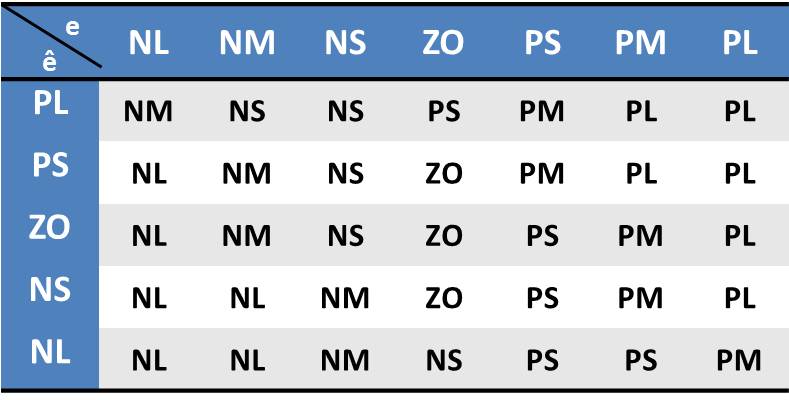}
  	  \end{center}
	\end{table}

   \begin{figure}[thpb]
     \centering
     \includegraphics[width=4cm,height=3cm]{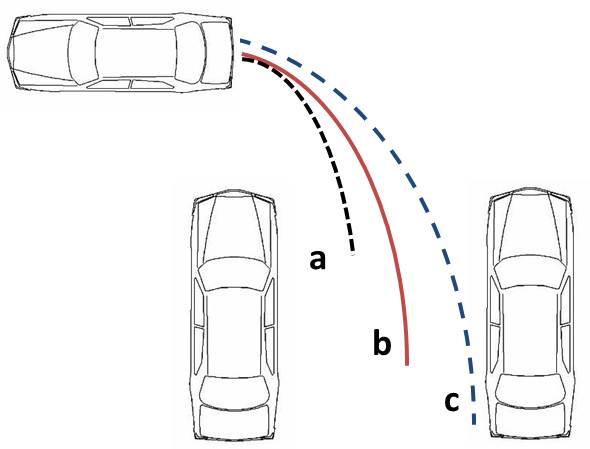}
     \caption{Turning Trajectories for Vehicles with Different Lengths}
     \label{figurelabel}
   \end{figure}

One input to the SFC is the angle difference (e) between set point and current angle while the other is the current turning radius ($r_1$). The current turning radius can be estimated as the ratio between linear velocity (v) and angular velocity ($\omega$) and normalized around zero as in Eq (1).

	\begin{equation}
		r_1=k_1\times \frac{v}{\omega}-1,	\text{ $k_1$ is the normalization coefficient}
	\end{equation}

$k_1$ is determined by the average turning radius of the entire trajectory. Therefore in ideal situation, the normalized radius should be negative at the start, zero in the middle and positive in the end. The output is an incremental voltage signal ($u_2$) sent to servo for steering angle control. It is added on top of the output signal ($u_1$) from BFC. (Note that the magnitude of voltage output from SFC is much smaller than that of BFC, usually around one-tenth). The membership functions for normalized radius (input) and incremental servo voltage (output) are given in Fig.4.

	\begin{figure}[thpb]     
     \centering$
\begin{array}{cc}
    \includegraphics[width=3.5cm,height=2.3cm]{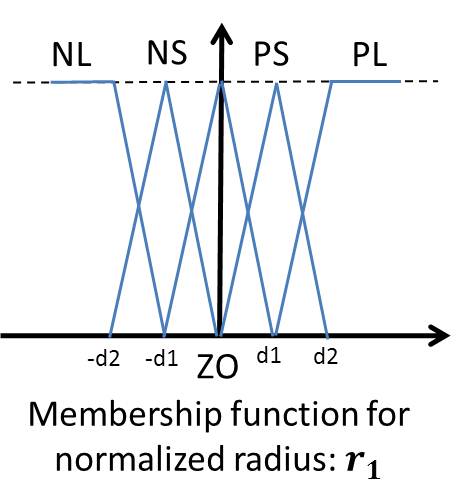}&
    \includegraphics[width=3.5cm,height=2.3cm]{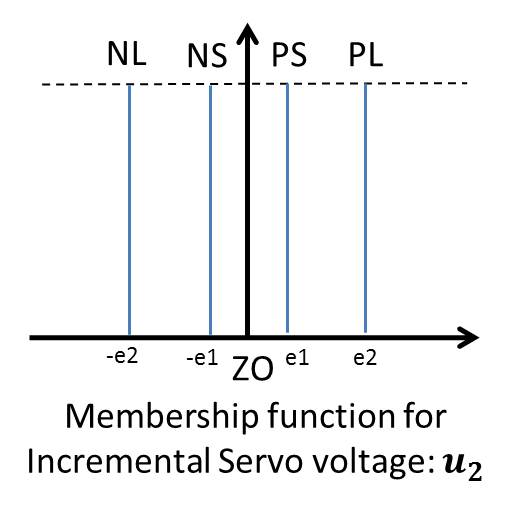}
\end{array}$
     \caption{SFC Inputs/Output in Steering Angle Control}    
   \end{figure}

Take vertical parking as an example, where the vehicle is trying to turn $90^\circ$ in the clockwise direction as in Fig.3. The control decision is made based on the following reasoning. At the beginning of turning, if the normalized radius is negative small, no modification is required and output voltage is zero. If it is negative large, the vehicle must be shorter than average, hence the steering angle should be reduced. A small positive voltage signal will be send to the servo to reduce the steering angle. If it is zero or positive, a small negative voltage should be sent so that the steering angle is increased. \footnote{Throughout the paper, it is assumed that turning in the clockwise direction is positive.} Expressed in IF-THEN rule format:

IF e is NL AND $r_1$ is NL, $u_2$ is PS;

IF e is NL AND $r_1$ is NS, $u_2$ is ZO;

IF e is NL AND $r_1$ is ZO, $u_2$ is NS;

IF e is NL AND $r_1$ is PS, $u_2$ is NL;

Table II summarizes the complete set of IF-THEN rules.

	\begin{table}
	  \caption{IF-THEN Rules for SFC in Steering Angle Control}
  	  \begin{center}
   	  \includegraphics[width=6cm,height=3cm]{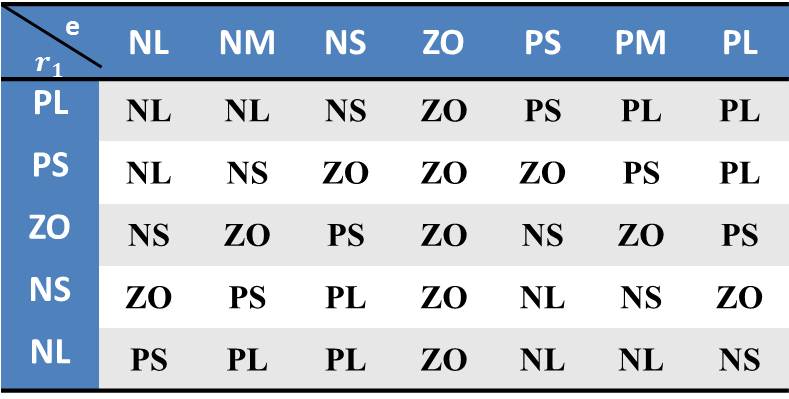}
  	  \end{center}
	\end{table}


\subsection{Speed Control}

   \begin{figure}[thpb]
     \centering
     \includegraphics[width=6.5cm,height=4cm]{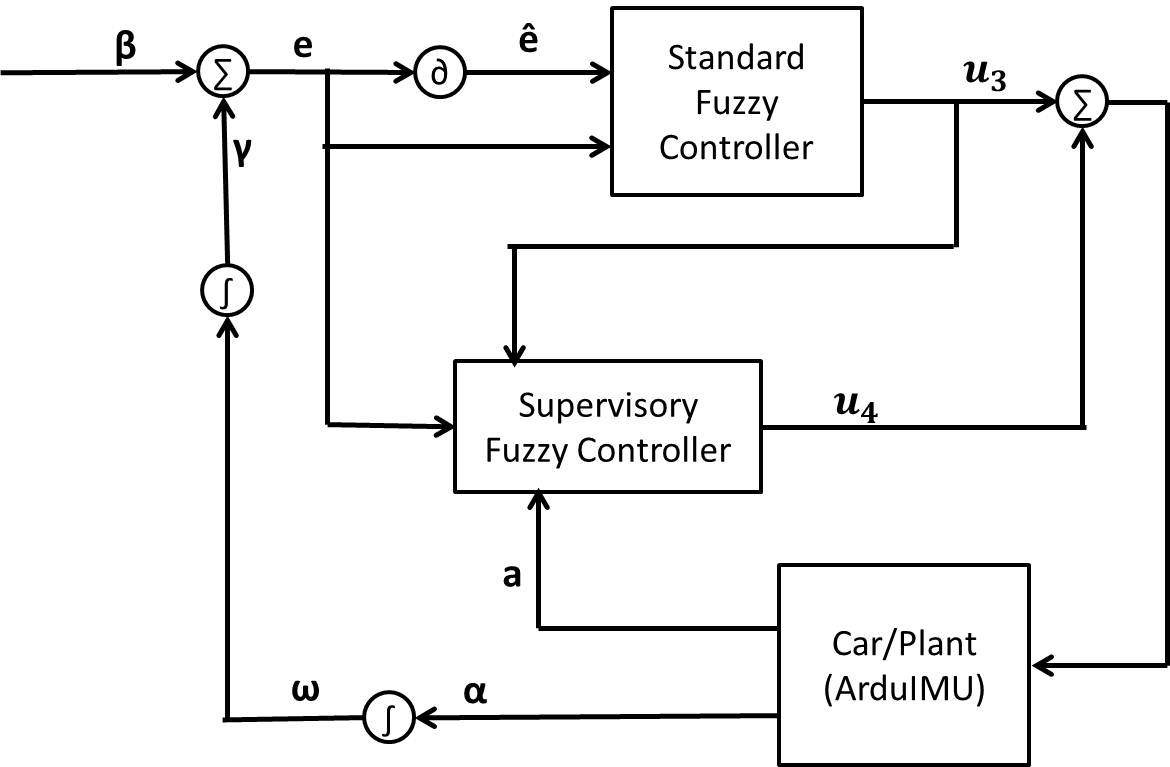}
     \caption{Speed Control}
     \label{figurelabel}
   \end{figure}

Fig.5 illustrates the control logic for the other path, i.e. speed control during turning. The speed is governed by the same principle and the inputs of BFC remain the same as in steering angle control, angle error e (between current angle and set point) and the change in that error $\hat{e}$. However, the case can be simplified since only magnitudes of the inputs are important. The output is a signal $u_3$ that sets the Pulse-Width-Modulation (PWM) duty cycle. The PWM duty cycle governs the power input to motor, hence the speed. The larger the duty cycle, the larger the speed. The membership functions for inputs and output are shown in Fig.6, while Table III summarises the IF-THEN rules.

	\begin{figure}[thpb]     
     \centering$
\begin{array}{ccc}
    \includegraphics[width=2.5cm,height=2cm]{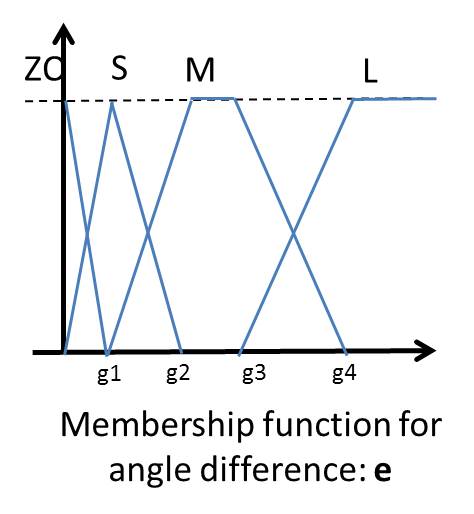}&
    \includegraphics[width=2.5cm,height=2cm]{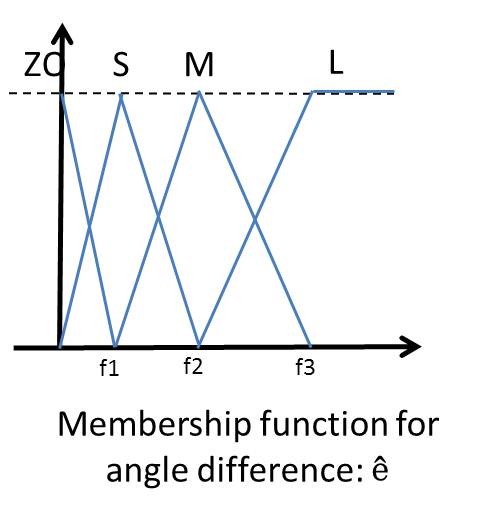}&
	\includegraphics[width=2.5cm,height=2cm]{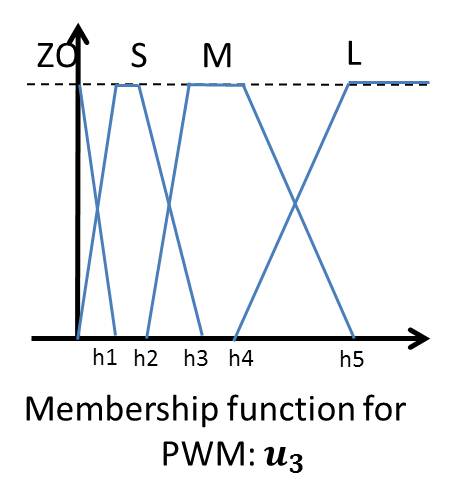}
\end{array}$
     \caption{BFC Inputs/Output in Speed Control}    
   \end{figure}

	\begin{table}
	  \caption{IF-THEN Rules for BFC in Speed Control}
  	  \begin{center}
   	  \includegraphics[width=4cm,height=2.8cm]{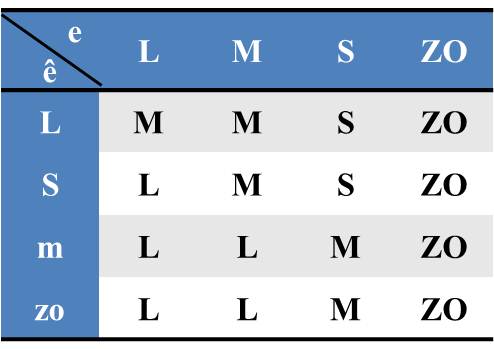}
  	  \end{center}
	\end{table}

Actual speed under same PWM may vary due to friction. Therefore, a SFC is designed to adjust PWM duty cycle so that the vehicle maintains a constant speed in different environments. If the friction is fixed, the ratio between velocity and PWM duty cycle should be fixed as well (as long as the velocity is stable). With larger friction, the ratio is smaller and vice versa. Hence one of the inputs to the SFC is given by Eq (2).

	\begin{equation}
		r_2=k_2\times \frac{v}{u_3}-1,	\text{ $k_2$ is the normalization coefficient}
	\end{equation}   
   
where $u_3$ is the output signal from BFC. The coefficient $k_2$ is defined such that $r_2$ equals zero when there is medium friction. Another input is the linear acceleration (a) measured by IMU. The output signal $u_4$ is the incremental signal to be added on top of the output signal of BFC. (Note that the magnitude of $u_4$ is much smaller compared to $u_3$, usually around one-tenth) The sum of $u_3$ and $u_4$ determines the duty cycle of the PWM controlling motor speed. The membership functions of the inputs/output are given in Fig.7. 

	\begin{figure}[thpb]     
     \centering$
\begin{array}{ccc}
    \includegraphics[width=2.7cm,height=2cm]{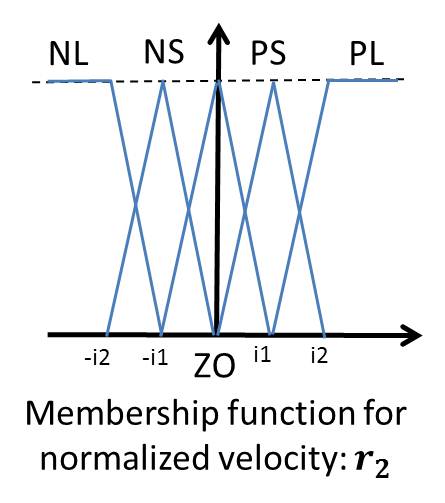}&
    \includegraphics[width=2.7cm,height=2cm]{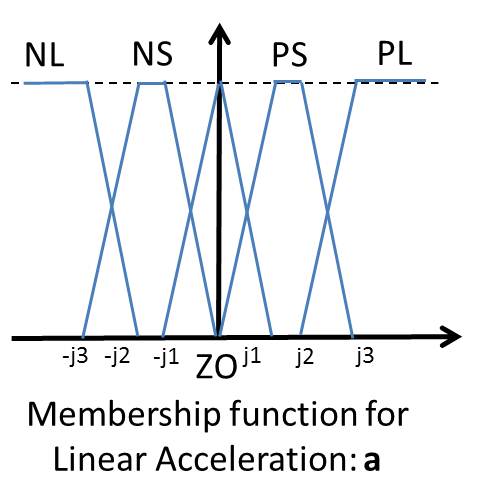}&
	\includegraphics[width=2.5cm,height=2cm]{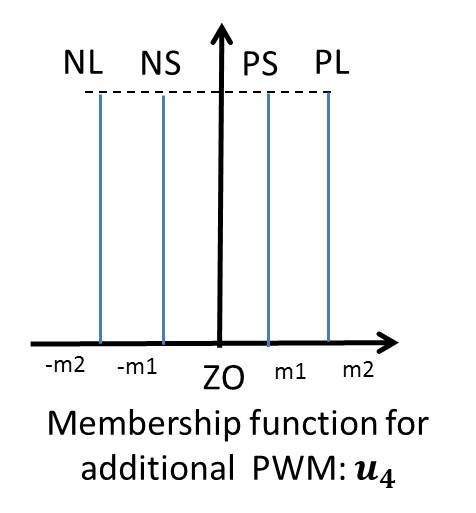}
	\end{array}$
     \caption{SFC Inputs/Output in Speed Control}    
   \end{figure}

Here are two examples of the SFC IF-THEN rules.

IF $r_2$ is NS AND a is ZO, $u_4$ is PS;

IF $r_2$ is NS AND a is PS, $u_4$ is ZO;

A negative small $r_2$ indicates that the friction is slightly larger than normal, hence the velocity does not reach the desired level. If the current acceleration is zero, the velocity has already reached stable state. Therefore, the PWM duty cycle should be increased to counteract additional friction. The SFC output $u_4$ is a positive small signal. However, if the acceleration is a small positive value, the implication is that velocity will be further increased. The velocity has not reached the stable value but it is moving in the correct direction. No modification is required at this stage. The system will evaluate the readings at next sampling time and adjust the SFC output accordingly. The complete IF-THEN rules are summarized in Table IV. 
   
   	\begin{table}
	  \caption{IF-THEN Rules for SFC in Speed Control}
  	  \begin{center}
   	  \includegraphics[width=4.5cm,height=3cm]{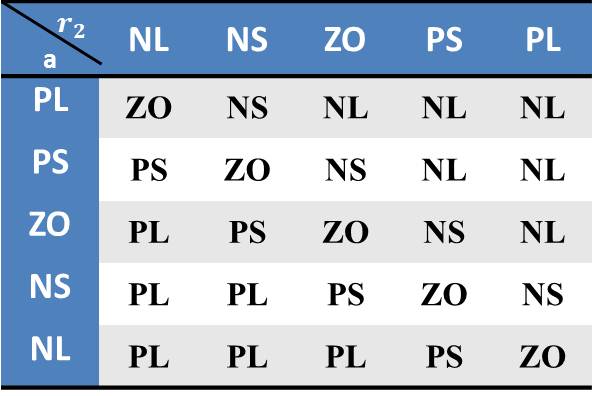}
  	  \end{center}
	\end{table}

\section{Posture Stabilization}

   \begin{figure}[thpb]
     \centering
     \includegraphics[width=7cm,height=4cm]{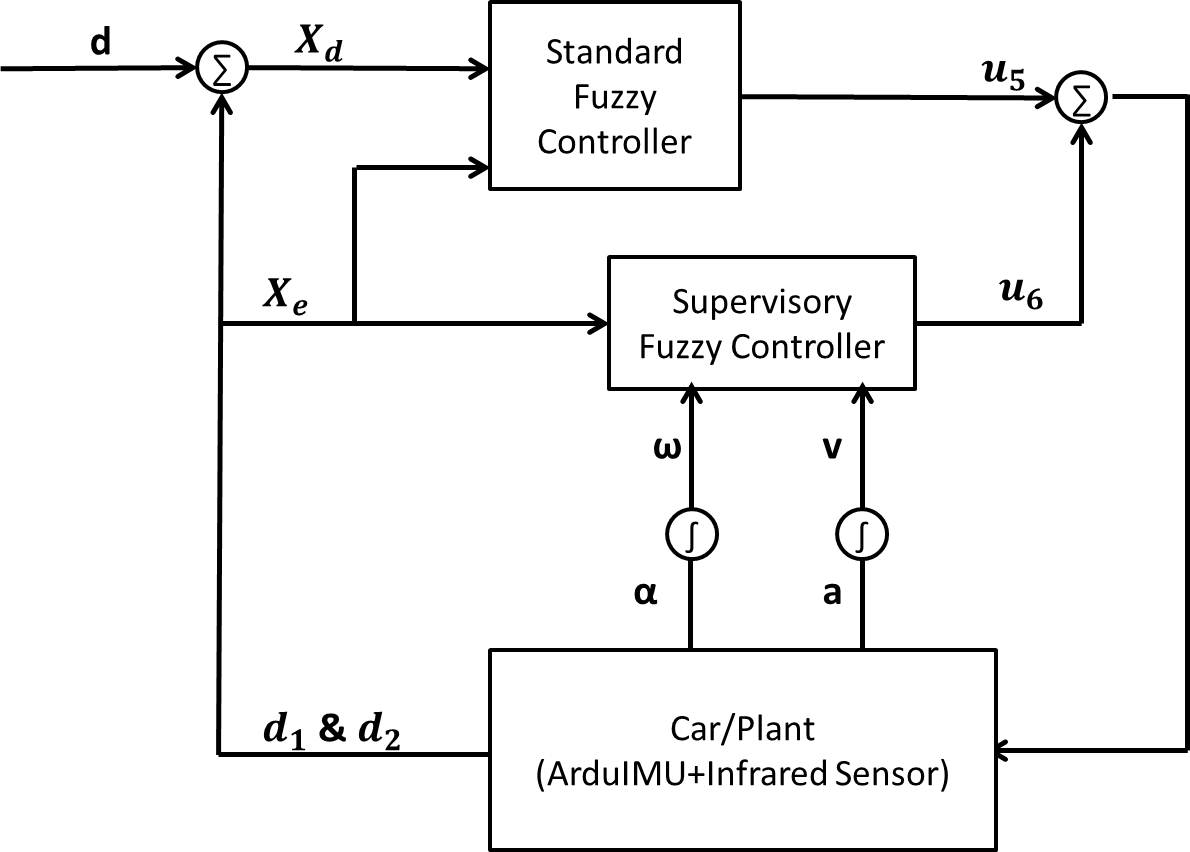}
     \caption{Posture Stabilization}
     \label{figurelabel}
   \end{figure}

In the previous FBOS design, posture stabilization is based on measurements from two infrared sensors located at the right side of the car, one at the front ($d_1$) and one at the back ($d_2$). The FBOS controls the car to move in a straight line while keeping a safe distance away from the wall by adjusting the steering angle of the front wheels. One of the inputs is the distance measurement of the sensor located at the front right-side of the car ($X_d$). The other input is the difference between the readings of the two sensors located at the right side ($X_e = d_1-d_2$). The output signal is the servo voltage ($u_5$) to control steering angle. The inputs/output membership functions are presented in Fig.8 while the IF-THEN rules are summarized in Table V.

	\begin{figure}[thpb]     
     \centering$
\begin{array}{cc}
    \includegraphics[width=2.7cm,height=2cm]{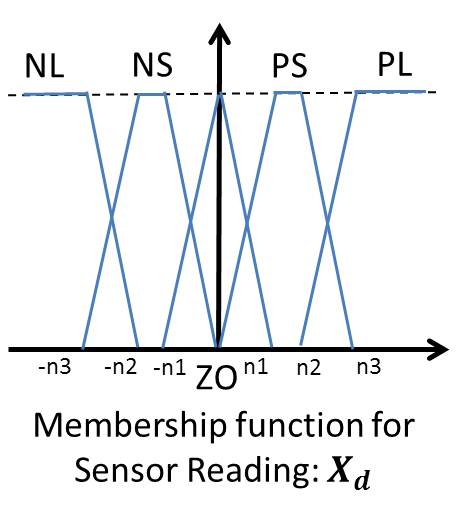}&
    \includegraphics[width=2.7cm,height=2cm]{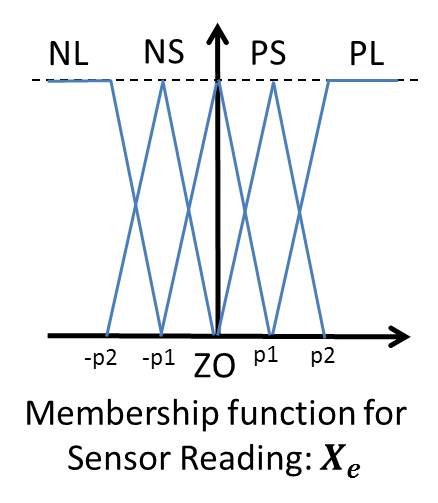}
	\includegraphics[width=2.7cm,height=2cm]{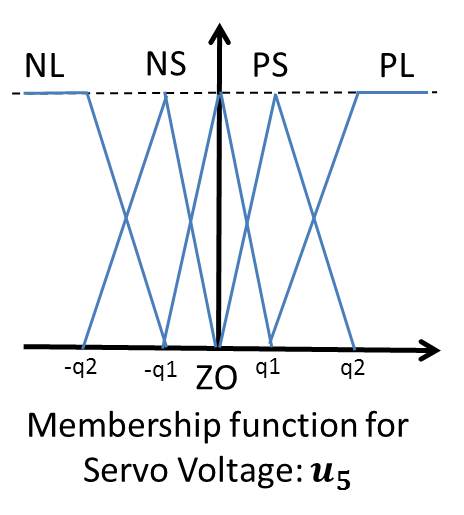}
\end{array}$
     \caption{BFC Inputs/Output in Posture Stabilization}    
   \end{figure}

The control rule is designed based on the assumption that a large $X_e$ indicates a large angular deviation from the forward-direction. However this is not always true. For a longer vehicle, a small angular deviation may result in a large $X_e$ since the two sensors are located further apart. On the contrary, a short vehicle may have a large angular deviation but a moderate $X_e$ due to the closeness of two sensors. Therefore, vehicle length must be taken into consideration for more effective posture stabilization. 

   	\begin{table}
	  \caption{IF-THEN Rules for BFC in Posture Stabilization}
  	  \begin{center}
   	  \includegraphics[width=4.5cm,height=3cm]{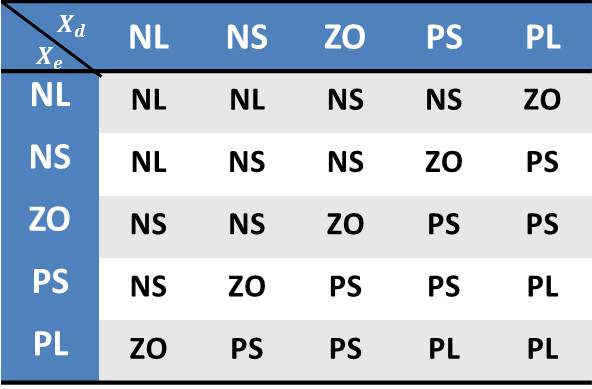}
  	  \end{center}
	\end{table}

Similar as in the case of steering angle control, a Hybrid Fuzzy Controller (HFC) is designed to improve the performance. The previous fuzzy logic controller can be treated as the Base Fuzzy Controller (BFC). A Supervisory Fuzzy Controller (SFC) is designed to adjust for the variation in vehicle length. The control diagram is illustrated in Fig.9.
   
	\begin{figure}[thpb]     
     \centering$
     \begin{array}{cc}
    \includegraphics[width=3.2cm,height=2cm]{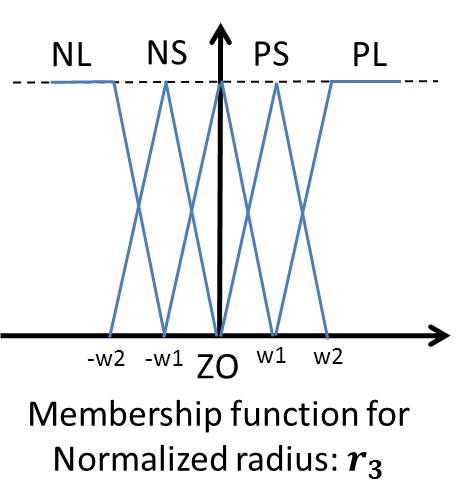}&
    \includegraphics[width=3.2cm,height=2cm]{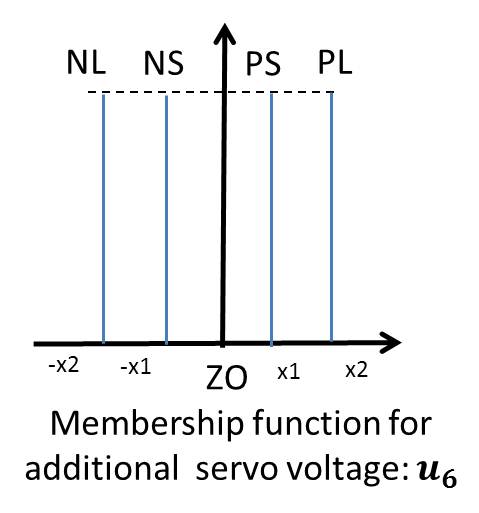}
\end{array}$
     \caption{SFC Input/Output in Posture Stabilization}    
   \end{figure}

Consider the case where a long vehicle has a small negative angular deviation. Since $X_e$ can be positive large in this case, the original control signal from BFC is positive large. With a large steering angle of the front wheel, $X_e$ changes quickly. As a result, there might be some unnecessary oscillation before the car can move forward steadily. As discussed before, longer vehicle has a larger turning radius if the front-wheel steering angle is the same. Therefore, the estimated turning radius can be taken as one of the inputs to the SFC. If both $X_e$ and estimated radius are large, the current vehicle has a larger size. The angular deviation is not large and the original steering angle can be reduced. The output signal of SFC is thus the incremental voltage signal sent to servo for steering angle adjustment ($u_6$). The estimated radius ($r_3$) is normalized by Eq (3).

	\begin{equation}
		r_3=k_3\times \frac{v}{\omega}-1,	\text{$k_3$ is the normalization coefficient}
	\end{equation}

Membership function for $X_e$ is the same as that in BFC, membership functions for $r_3$ and $u_6$ are shown in Fig.10. Table VI summarizes the IF-THEN rules of the SFC.
	\begin{table}
	  \caption{IF-THEN Rules for SFC in Posture Stabilization}
  	  \begin{center}
   	  \includegraphics[width=5cm,height=3cm]{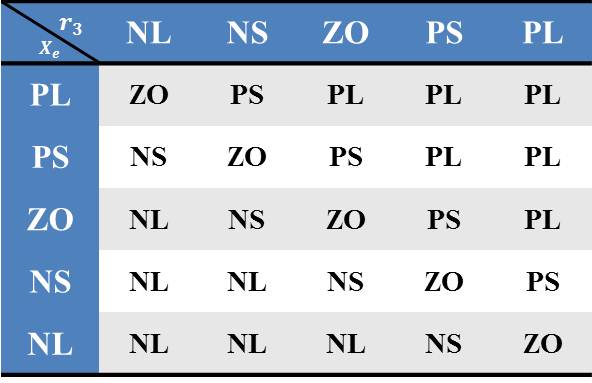}
  	  \end{center}
	\end{table}

\section{Simulation and Implementation}

Several experiments are simulated in MATLAB to verify the robustness and reliability of newly designed Hybrid Fuzzy Controllers (HFCs) by comparing the responses under HFC and original FBOS. The experiments are designed using Control Variable Method, meaning that only one disturbance is introduced at a time. The purpose is to evaluate the performance of individual HFC. The parameters in each fuzzy rule base are configured through trial before actual experiments. 
Snapshot pictures are extracted from simulation at a frequency of 0.5 frame per second. Performance of different controllers can be compared using the sequential images of the parking process. In addition, steering angle of the front wheels is illustrated for reference.

Both parallel and vertical parking are controlled under the same fundamental principles. Only the test results of parallel parking are demonstrated here due to space constraint. 

The first experiment is designed to test the robustness of improved HFC when the vehicle length varies. A comparison test is done by using the original FBOS. In order to make the simulation closer to real-life situations, the dimension of the long vehicle is set to one-fourteenth of a normal bus, i.e. 500mm in length and 180mm in width. Fig.11 presents the sequential images extracted from simulated experiment of long vehicle parallel parking controlled by original FBOS. Comparatively, Fig.12 demonstrates the parallel parking process of long vehicle controlled by improved HFC. 

   \begin{figure}[thpb]
     \centering
     \includegraphics[width=7cm,height=1.8cm]{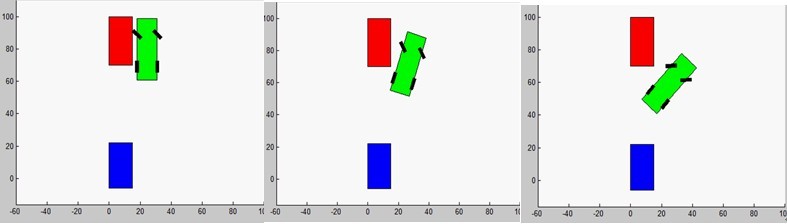}
     \includegraphics[width=7cm,height=1.8cm]{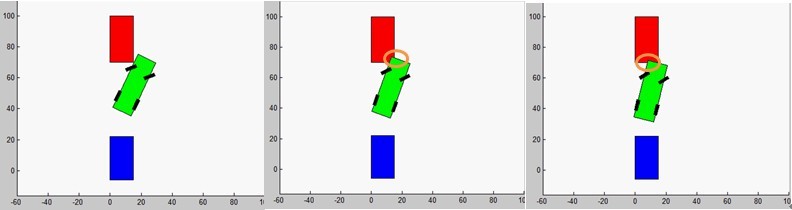}
     \caption{Long vehicle parallel parking under the control of FBOS}
     \label{figurelabel}
   \end{figure}

In Fig.11, it is observed in the last few pictures that a collision occurs between the vehicle and the one parked in neighbouring slot. The result shows that steering configuration tuned for a small car does not guarantee same satisfactory performance when applied to a longer vehicle, confirming the previous assumption that longer vehicles usually require larger turning radius. In Fig.12, the issue is resolved with the improved HFC. Based on real-time measurements from sensors and IMU, the vehicle adjusts the steering angles to adapt to variation in length, thus avoiding significant deviation from the optimal turning trajectory.

   \begin{figure}[thpb]
     \centering
     \includegraphics[width=7cm,height=1.8cm]{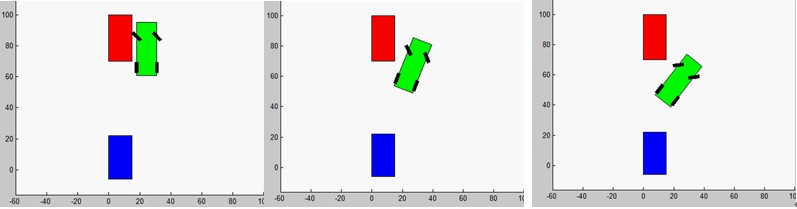}
     \includegraphics[width=7cm,height=1.8cm]{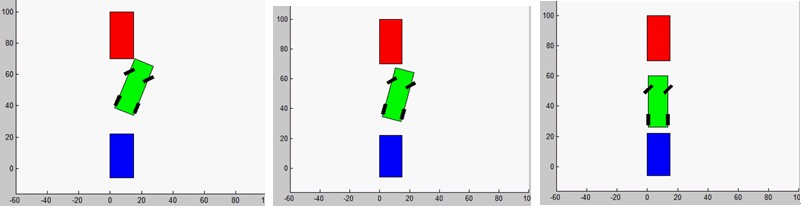}
     \caption{Long vehicle parallel parking under the control of AFC}
     \label{figurelabel}
   \end{figure}

The second experiment is designed in a similar way to test the robustness of improved HFC with different ground conditions. Simulated experiments are conducted to compare the two controllers, given same changes (increased friction) in parking environment. The friction simulated here follows Hook's law, f=$\mu$mg, where $\mu$ is the friction coefficient. Increased friction counteracts part of the motor power, thus slowing down the entire parking process. 

Fig.13 presents the captured images of parking process under the control of original FBOS. 
Since the images are captured every two seconds, parking time consumed is around 12 seconds since only the last image indicates complete parking. Fig.14 illustrates the experiment result using improved HFC. The entire parking process ends at the fifth image, where the total time required is only around 10 seconds. Comparing the time cost of two parking processes, it can be easily determined that HFC has superior performance than original FBOS.

   \begin{figure}[thpb]
     \centering
     \includegraphics[width=7cm,height=1.8cm]{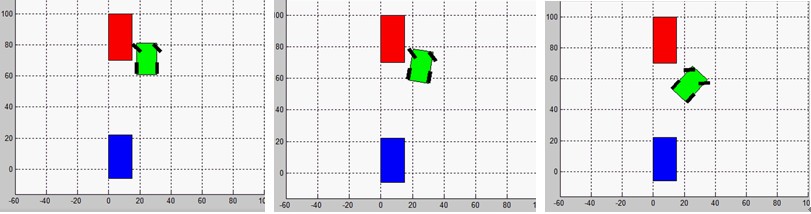}
     \includegraphics[width=7cm,height=1.8cm]{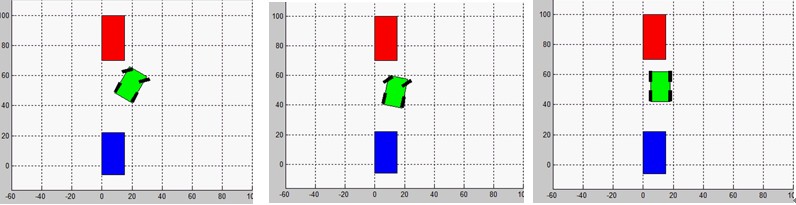}
     \caption{Coarse ground parallel parking under the control of FBOS}
     \label{figurelabel}
   \end{figure}

   \begin{figure}[thpb]
     \centering
     \includegraphics[width=7cm,height=1.8cm]{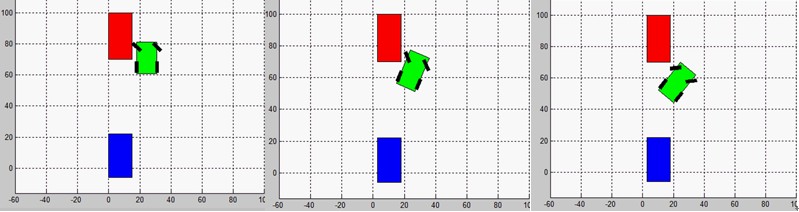}
     \includegraphics[width=7cm,height=1.8cm]{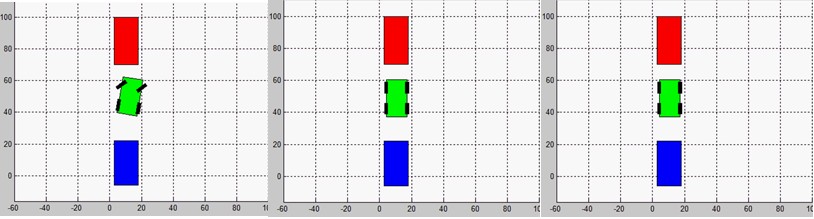}
     \caption{Coarse ground parallel parking under the control of AFC}
     \label{figurelabel}
   \end{figure}

\section{Conclusions and Future Work}

In this paper, an intelligent autonomous parking system with Hybrid Fuzzy Controller (HFC) is proposed. Vehicles equipped with the system are able to perform slot detection and auto-parking in either parallel or vertical parking mode. A major improvement from previous work is the design and implementation of HFCs for two critical parking steps, turning control and posture stabilization. In turning control, two separate control paths, steering angle control and speed control, work simultaneously to achieve better performance. Both paths use the proposed HFC, consisting of a Base Fuzzy Controller (BFC) and a Supervisory Fuzzy Controller (SFC). In posture stabilization, the proposed HFC is also used to improve robustness. The optimized controllers ensure a smooth and efficient parking process even when there are variations in vehicle length and ground conditions. 

The deployment of HFC improves the system performance, yet at the same time opens up more opportunities for future research. The current SFC only modifies BFC's output. Future work can look into the possibilities of dynamic fuzzy rule base. There is great potential to make the system even more robust and efficient. However, the complications coming along should be addressed properly. With more advanced control rules, the complexity of implementation increases exponentially. Whether the improvement in performance is able to justify for the increased cost should be carefully analysed since practical application is one of the main objectives in developing this auto-parking system.      


\end{document}